### ARTICLE

# Disorder scattering in classical flat channel transport of particles between twisted magnetic square patterns

Anna M. E. B. Rossi  [1], Adrian Ernst  [1], Magdalena Dörfler[1] & Thomas M. Fischer  [1✉]

We measure the trajectories of macroscopic magnetic particles pulled against gravity between twisted alternating magnetic square patterns in a superposed homogeneous magnetic field normal to both patterns. The two patterns are built from a set of magentic cubes having a distribution of magnetization. The magnetic potential between the patterns is a sum of three contributions: two being periodic on two lattices with different magnitude and orientation, and the third random contribution arising from the distribution of magnetization of the cubes. As one varies the twist angle between the two patterns each time the twist angle coincides with a magic twist angle one of the two periodic lattices becomes a sublattice of the other lattice. Simulations of particles moving through patterns with a precise cube magnetization produce pronounced mobility peaks near magic twist angles that are associated with flat channels. Weak random fluctuations of the cube magnetization in the experiment and the simulations cause enhanced random disorder of the potential and reduce the mobility by scattering particles into the interior of the twisted Wigner Seitz cells. The mobility undergoes an Anderson transition from magic to generic behavior as the magnetization disorder increases beyond half of a percent of the cube magnetization.

[1] Institute of Physics, Universität Bayreuth, 95440 Bayreuth, Germany. ✉email: Thomas.Fischer@uni-bayreuth.de

Electric currents in twisted hexagonal structures such as twisted bilayer graphene[1–7], twisted bilayers of the transition metal dichalcogenide family[8], the transport of waves in twisted photonic[9–13] or acoustic[14–17] crystals or in vortex lattices[18] are subject of intense research, in part because they might be model systems for a better understanding of the emergence of non-conventional superconductivity[19–25] and ferromagnetism[26,27]. A bilayer consisting of a stack of two identical monolayers with the second monolayer rotated by a twist angle relative to the first monolayer generically creates a quasi-crystalline Moiré pattern. The resulting Moiré pattern potential is a function of the local relative and the total shift of each nearest monolayer unit cell center from the chosen location. Both the relative and the total shift are periodic functions on two different lattices, the larger $\mathcal{Q}$-lattice, and the smaller $\mathcal{P}$-lattice. For specific nongeneric, magic twist angles, the $\mathcal{Q}$-lattice is a sublattice of the $\mathcal{P}$-lattice. Thus, magic Moiré patterns are periodic on the $\mathcal{Q}$-lattice with a Moiré Wigner Seitz cell ($\mathcal{W}_{\mathcal{DV}} = \mathcal{W}_{\mathcal{Q}}$) extending over mesoscopic classical distances. The Brillouin zone of the monolayer is folded into a mini Brillouin zone ($\mathcal{MBZ}$) of the magic bilayer Moiré crystal, and the electronic band structure of a monolayer splits into multiple mini bands and mini gaps. Amongst these mini bands, flat bands[20,21] with a very small bandwidth are responsible for the peculiar transport behavior[5,19,28]. One may tune the Fermi energy of an electronic twisted bilayer system through a flat band, and unconventional superconductivity seems to be most likely to occur at the energies associated with van Hove singularities of the band that lie in the middle of the flat bands. From the Bloch wave functions, we may define quasi-momentum as well as direct space densities of states. The local densities of state for energy at a flat band minimum in twisted bilayer graphene are localized at the $K$ and $K'$ points of the mini Brillouin zone and at the $AA$-stacking points of the Moiré Wigner Seitz cell and are thus localized in quasi-momentum and in direct space[5,28,29]. Both densities of state are also localized for energy near the flat band maximum, and they peak at three stacking locations in the $\Gamma M$ line in the mini Brillouin zone and to the $AB$ and $BA$ stacking locations in the Moiré Wigner Seitz cell. At the van Hove singularity in the middle of the flat band, however, the density of states is delocalized both in quasi-momentum and in real space with maximal density on top of lemniscates that are constant energy levels connecting all boundaries of the mini Brillouin zone, respectively the Moiré Wigner Seitz cell that runs through the saddle points of the energy. The degenerate states at the energy of the van Hove singularity thus create the extended density of states, contributing to the unconventional superconductivity of twisted electronic systems.

Twisted van der Waals Moiré bilayer systems have also been created with micromagnetic monolayers[30–33]. Macroscopic twisted systems share all symmetry properties with their microscopic analogs. They can thus be used to unravel and visualize transport modes that are difficult to access at the microscopic level. Extended states are vulnerable to disorder[29,34–38] and can be localized via the Anderson localization[29,39] as observed in photonic wave systems[40] including Aubry André Moiré systems[12] in ultracold atoms[41], Bose–Einstein condensates[42], as well as in ultrasonic waves of soft matter elastic media[43]. In this work, we show that the relocalization of extended particle states analog to the Anderson localization can obscure the peculiarities of classical twisted magnetic transport systems.

## Results

**Experimental setup.** In previous work, we have shown that topologically nontrivial transport appears in colloidal[44–51] and in macroscopic[52,53] magnetic particle systems that are subject to dissipative dynamics. Here, we use macroscopic magnetic parti-
cles immersed in water to drive these particles through the magnetic potential created between two shifted and oppositely rotated magnetic square patterns. Between exactly periodic twisted patterns, colloidal particles[54] move along a zig–zag course in flat channels with potential energies well below the average potential energy level of the untwisted patterns. The channels have infinite lengths for magic twist angles between the patterns if a convenient relative shift of both patterns is chosen. For non-magic angles, the relative shift between both patterns repeats itself on the $\mathcal{Q}$-lattice. However, the total shift that varies on the $\mathcal{P}$-lattice lets the potential differ between neighboring pseudo-Wigner Seitz cells of the $\mathcal{Q}$-lattice. For this reason, the zig–zag channels of non-magically twisted patterns have finite lengths. The mobility of colloidal magnetic particles predicted with computer simulations, therefore, is a discontinuous function of the twist angle with pronounced peaks at the magic angles[54].

In a macroscopic experimental situation, slight local variations of the magnetization of the elements of the twisted patterns are unavoidable. These variations cause large relative local deviations in the normal component of the mid-plane magnetic field that scatters experimental particles out of the flat channels.

In Fig. 1, we show schematics of our experimental setup. It consists of two magnetic square patterns of $2 \times 2523$ alternating (gray and white) magnetized N45-$Nd_2Fe_{14}B$ cubes of side length 2.03 mm and magnetization $\mu_0 M_s = 1.345 T \pm 0.025 T$ arranged in a plane and mechanically stabilized within an epoxy resin. The magnetization of the cubes has a distribution with a relative width of 2%. The untwisted pattern is oriented with the sum of the two (yellow) primitive unit vectors $\boldsymbol{a}_1 + \boldsymbol{a}_2 = -a\sqrt{2}\boldsymbol{e}_z$ of the individual patterns along the direction of gravity, where $a$ is the lattice constant, and $\boldsymbol{e}_z$ is a unit vector that points against gravity. The left pattern and right pattern planes are parallel to each other at a separation of $d = 6.1$ mm. We define the center of the Wigner Seitz cell of both patterns to be the $C_4$ symmetric point of the right magnetized (white) cube. An axis of rotation normal to both patterns (upper green arrow) allows both patterns to be rotated individually about this externally defined axis near the top edge of the pattern. The axis of rotation of both patterns can be set directly, and the angle of rotation is adjusted with a laser. We place the origin of our 2D midplane coordinate system right into this external axis of rotation. Before we rotate the patterns, we shift them by two shift vectors $s_l(0)$ and $s_r(0)$, such that the axis of rotation runs through an $S_4$-symmetric point of the left pattern and through a $C_4$-symmetric point of a right magnetized cube of the right pattern. Hence, the left pattern has its center of the Wigner Seitz cell shifted by half of the first primitive unit vector $s_l(0) = a_1/2$ while the center of the Wigner Seitz cell of the right pattern is unshifted $s_r(0) = 0$. Both patterns are thus relatively shifted by half of the first primitive unit vector $a_1/2$ such that all $C_4$ symmetric points of the left pattern are opposed by $S_4$ symmetric points of the left pattern. Without twist, this generates a stripe-patterned magnetic field with stripes along the $a_1$-(zig)-direction normal to the $a_2$-(zag)-direction and at an angle of $\pi/4$ with respect to the direction of gravity.

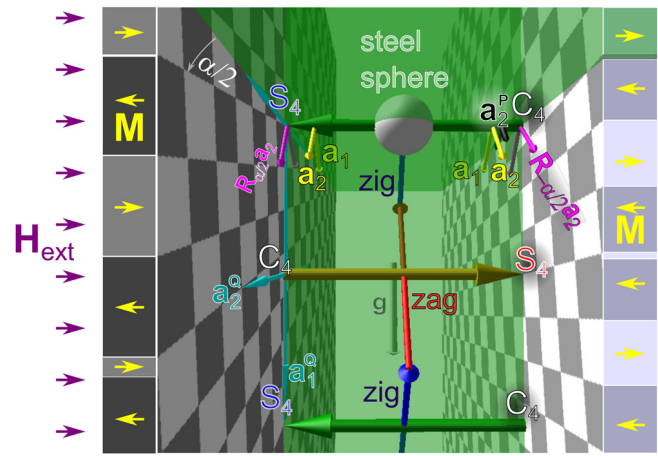

**Fig. 1 Scheme of the setup.** A steel sphere is confined to the mid-plane between two twisted magnetic patterns. Both patterns are shifted with respect to each other and rotated by $\pm \alpha/2$ around an axis perpendicular to the patterns (upper green arrow). Both magnetic patterns are square patterns with left (gray) and right (white) magnetized cubes. Primitive unit vectors of the nonrotated pattern (yellow) of the rotated patterns (magenta) as well as of the $\mathcal{P}$-lattice (black) and the $\mathcal{Q}$-lattice (cyan) are shown. The direction of the gravitational acceleration $\boldsymbol{g}$ is indicated by the arrow in the back. An external magnetic field $\boldsymbol{H}_{ext}$ (purple) parallel to the magnetization of the white cubes is superposed to the magnetic fields from both patterns. In a perfectly ordered system the steel sphere is lifted along a flat channel following the maximal overlap with the white cubes on both sides. This zig-zag path between both patterns does not stop if the twist angle $\alpha$ is magic $\alpha_m(k) = 2\arctan(1/(2k+1))$. The figure shows a magic, $\alpha_m(3) = 2\arctan(1/7)$, non generic situation.

We then rotate both patterns by an angle $\pm\alpha/2$ in opposite directions around the external axis. The twist generically destroys the opposition of $C_4$ and $S_4$ symmetry points for all points but those lying on both the $\mathcal{P}$- and the $\mathcal{Q}$-lattice to be explained further below (primitive unit vectors of the $\mathcal{P}$- and the $\mathcal{Q}$-lattice are shown in black and cyan in Fig. 1). Both patterns create a magnetic field $H_p(r)$ of magnitude $\mu_0 H_p \approx 3$ mT in the midplane between both patterns. A water tank with an inner distance of 2 mm is placed between both patterns to dampen the movement of a steel sphere placed inside. To the left of both patterns, we place a set of Helmholtz coils, creating a homogeneous external magnetic field $H_{ext}$ (purple) of magnitude $\mu_0 H_{ext} \approx 29$ mT that points toward the right, i.e., normal to both patterns and penetrates the entire space between both patterns. The 1.9 mm diameter steel sphere with a nearly linear magnetization curve $M(H)$ for $\mu_0 H < 30$ mT is glued to the middle of a hair fiber removed from the first author. At the lower end of the hair, we attach a weight of $m_W = 50$ mg. Initially, we release the steel sphere and the weight such that the steel sphere hangs near the lower end of the pattern below the axis of rotation. The length of the hair ensures that the nonmagnetic weight below the nonvisible steel sphere remains visible. We can infer the position of the steel sphere from the position of the weight. Above the pattern, the hair passes an eyelet. We lift the upper end of the hair with a speed of $v_{lift} = 10^{-3}$ m s$^{-1}$. The velocity is slow enough for the weight to adiabatically follow the motion of the steel sphere with the direction of the hair between the weight and steel sphere pointing along the direction of gravity. The position of the steel sphere is confined to the midplane by the walls of the water tank. When lifted against gravity between two perfect patterns, the steel sphere is expected to follow a zig–zag course in the mid-plane between both patterns and therefore does not remain right below the eyelet.

The steel sphere of effective susceptibility $\chi_{eff}$ and volume $V$ acquires a fixed magnetic moment $\mathbf{m} \approx \chi_{eff} V \mu_0 H_{ext}$ dominated by the external field. The potential energy of the steel sphere in the midplane between both patterns can be written as[54]

$$
\begin{aligned}
U &\approx \boldsymbol{m} \cdot \boldsymbol{H}_p(\boldsymbol{r}) \\
&\propto \underbrace{\sum_{i=1}^{4} \left[ \cos(\boldsymbol{k}^i \cdot \boldsymbol{s}_l(\boldsymbol{r})) + \cos(\boldsymbol{k}^i \cdot \boldsymbol{s}_r(\boldsymbol{r})) \right]}_{\text{quasi-periodic}} \\
&+ \underbrace{e^{kd/2} \left( 1 - \frac{d/2}{\sqrt{\xi^2 + d^2/4}} \right) \frac{\delta M_z(\boldsymbol{r})}{\bar{M}_z}}_{\text{random}}
\end{aligned}
\tag{1}
$$

The potential energy is proportional to the sum of a quasi-periodic and a random contribution.

**Quasi-periodic potential contribution.** Let us discuss the quasi-periodic contribution first: The $\boldsymbol{k}^i = R_{\pi/2}^{i-1} \cdot \boldsymbol{k}^1$ in the first term of Eq. (1) are the four ($i = 1 .. 4$) primitive reciprocal unit vectors of the non-rotated pattern, and $R_{\pi/2}$ and $R_{\pm\alpha/2}$ are rotation matrices by the angles $\pi/2$ and $\pm\alpha/2$. The shift vectors $\boldsymbol{s}_{l,r}(\boldsymbol{r}) = R_{\pm\alpha/2}^{-1} \cdot (\boldsymbol{r} - \boldsymbol{r}_{center,lr}) \stackrel{\text{mod}\, \boldsymbol{a}_1}{=} \boldsymbol{s}_{l,r}(\boldsymbol{0}) + R_{\pm\alpha/2}^{-1} \cdot \boldsymbol{r}$ of the left and right pattern are the vectors from the nearest Wigner Seitz cell centers in each of the rotated patterns toward the position of interest $\boldsymbol{r}$, but rotated back into the unrotated pattern orientation. The abbreviation ($\bmod \boldsymbol{a}_1, \boldsymbol{a}_2$) above the equal sign indicates that the left and right sides of the equation are equal up to differences of integer multiples of the lattice vectors of the unrotated pattern. Both shift vectors map each position of interest $\boldsymbol{r}$ into the unrotated central single pattern Wigner Seitz cell $\mathcal{W}$. The primitive unit vectors of the $\mathcal{Q}$-lattice are $\boldsymbol{a}_i^{\mathcal{Q}} = (\boldsymbol{a}_i + \boldsymbol{a}_{i+1})/2 \sin(\alpha/2)$, where $\boldsymbol{a}_{i+1} = R_{\pi/2} \cdot \boldsymbol{a}_i$ are all primitive unit vectors of both patterns prior to rotation. The primitive unit vectors of the $\mathcal{P}$-lattice are $\boldsymbol{a}_i^{\mathcal{P}} = \boldsymbol{a}_i/2 \cos(\alpha/2)$. The primitive unit vectors of the $\mathcal{Q}$-lattice point along the direction of gravity and along the horizontal direction (see Fig. 2b)), while the $\mathcal{P}$-lattice is oriented the same way as the original unrotated lattices of the individual patterns (see Fig. 2a) or d)). We can express the two shift vectors as

$$
\begin{pmatrix} \boldsymbol{s}_l(\boldsymbol{r}) \\ \boldsymbol{s}_r(\boldsymbol{r}) \end{pmatrix} \stackrel{\text{mod}\, \boldsymbol{a}_1, \boldsymbol{a}_2}{=} \frac{1}{2} \left\{ \underbrace{(\boldsymbol{s}_l(\boldsymbol{r}) + \boldsymbol{s}_r(\boldsymbol{r})) \begin{pmatrix} 1 \\ 1 \end{pmatrix}}_{\text{periodic on the }\mathcal{P}\text{-lattice}} + \underbrace{(\boldsymbol{s}_l(\boldsymbol{r}) - \boldsymbol{s}_r(\boldsymbol{r})) \begin{pmatrix} 1 \\ -1 \end{pmatrix}}_{\text{periodic on the }\mathcal{Q}\text{-lattice}} \right\}
\tag{2}
$$

a sum of two functions that separately return to the same point in the Wigner Seitz cell $\mathcal{W}$ of the single pattern lattice if one advances the position by a primitive unit vector of the two different combined pattern lattices $\boldsymbol{r} \rightarrow \boldsymbol{r} + \boldsymbol{a}_i^{\mathcal{P},\mathcal{Q}}$.

The positions of negative interference between the rotated left and right pattern are those where the two rotated patterns have a relative shift at the corner of the individual non-rotated pattern Wigner Seitz cell $\boldsymbol{s}_l(\boldsymbol{r}) - \boldsymbol{s}_r(\boldsymbol{r}) = (\boldsymbol{a}_1 + \boldsymbol{a}_2)/2$. There are two such positions within each Wigner Seitz cell $\mathcal{W}_\mathcal{Q}$ of the $\mathcal{Q}$-lattice, namely $\boldsymbol{c}_{\pm}^{\mathcal{Q}} = \pm(\boldsymbol{a}_1^{\mathcal{Q}} - \boldsymbol{a}_2^{\mathcal{Q}})/4$ (see Fig. 2a). We call these positions the corners. The value of the quasi-periodic part of the magnetic potential of the particles at the corners vanishes, $U(\boldsymbol{c}_{\pm}^{\mathcal{Q}}) = 0$, and is right between the maximum and minimum potential value (a van Hove singularity of the potential energy density). The origin is located in the middle (upper green arrow in Fig. 1, green point in Fig. 2) of the zig-channel segment (blue arrow in Fig. 1 and blue in Fig. 2), where $\boldsymbol{s}_l(\boldsymbol{0}) = \boldsymbol{a}_1/2$ and $\boldsymbol{s}_r(\boldsymbol{0}) = \boldsymbol{0}$. The channel segment ends close to the nearest $\boldsymbol{c}_{\pm}^{\mathcal{Q}}$-corner and is (see Fig. 2a) or is not (see Fig. 2b, c) connected to a zag-segment (red arrow in Fig. 1). In the magic case, the zag-segment again connects while in the generic case, it does not connect to a further segment in zig-direction near the next $\boldsymbol{c}_{\pm}^{\mathcal{Q}}$-corner. For commensurate periods, the twist angle satisfies the condition $\alpha(q,p) = 2 \arctan(q/p)$ with $q, p \in \mathbb{Z}$ coprime numbers. For the magic angles $\alpha_m(k) = \alpha(q = 1, p = 2k + 1) = 2 \arctan(1/(2k+1))$, $k \in \mathbb{Z}/\{0\}$ the $\mathcal{Q}$-lattice is a sublattice of the $\mathcal{P}$-lattice and the entire channel continues its zig-zag path from one corner to the next. The zig–zag direction is along the direction of gravity for $\text{sign}[(-1)^k \alpha_m(k)] > 0$ and along the horizontal direction for $\mathrm{sign}[(-1)^k \alpha_m(k)] < 0$. For non magic but commensurate periods, the zig-zag trajectory must pass $c = 2q2^{(p-q)\mathrm{mod}\,2}$ different corners of the $q^2 2^{(p-q)\mathrm{mod}\,2}$ pseudo Wigner Seitz cells $\mathcal{W}_Q$ within one twisted Wigner Seitz cell $\mathcal{W}_{TW}$. Generically in such case, there is at least one corner that is poorly connected and, therefore, difficult to pass by a particle.

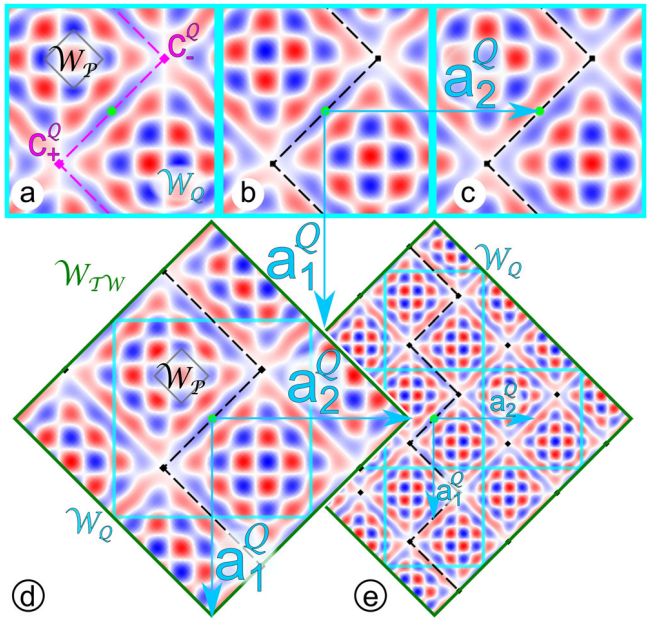

**Fig. 2 Disorder free mid-plane potentials.** (As seen when looking from the left pattern toward the right pattern) for different angles and shift vectors for a commensurate situation within the corresponding twisted Wigner Seitz cells $\mathcal{W}_{TW}$ of the potential. Red colors are maxima, and blue colors are minima of the potential. Panel **a**-**c** are potentials with the magic twist angle $\alpha_m(3) = 2\arctan(1/7)$ for **a** shift vectors $s_l(0) = a_1/2$ and $s_r(0) = 0$, **b** shift vectors $s_l(0) = a_1/4$ and $s_r(0) = -a_1/4$, **c** shift vectors $s_l(0) = (a_1 - a_2)/4$ and $s_r(0) = -(a_1 - a_2)/4$. Panel **a**-**c** show a situation where there exist connected [non-connected] flat channels between the corners $c_{\pm}^Q$. For the magic conditions, the twisted Wigner Seitz cell coincides with the Wigner Seitz cell of the $\mathcal{Q}$-lattice $\mathcal{W}_{TW} = W_Q$. The cyan primitive unit vectors $a_i^Q$ are shown in panels (**b**, **d**, **e**). In the commensurate cases in panels **d** ($\alpha(1,8) = 2\arctan(1/8)$) and **e** ($\alpha(3,22) = 2\arctan(3/22)$) the twisted Wigner Seitz cell (dark green) consists of 2, respectively 8 pseudo-Wigner Seitz cells (cyan lines) of the $\mathcal{Q}$-lattice that all differ from one another. In **a**, **d**, we also mark one pseudo-Wigner Seitz cell $\mathcal{W}_P$ of the $\mathcal{P}$-lattice (black).

Simulated particles driven with a weak constant force[54] follow this zig-zag path through the entire pattern if the twist angle is magic and if the zig-zag path is along the pulling direction against gravity. In this specific magical case, each of the Wigner Seitz cells of the $\mathcal{Q}$-lattice are equivalent. If the twist angle is not magic, the periodic part on the $\mathcal{P}$-lattice modulates the potential in a way that the potential in each pseudo-Wigner Seitz cell $\mathcal{W}_Q$ of the $\mathcal{Q}$-lattice differs from the previous one. The simulated particles, therefore, stop moving near one of the corners under non-magic twist conditions.

**Magnetization disorder potential.** We now discuss the effect of the random part of the potential in Eq. (1) that arises from the distribution of the magnetic cube magnetization. The magnetic pattern field $H_p(r)$ fulfills the Laplace equation and, therefore, periodic Fourier components to the field with the primitive reciprocal unit vectors $\mathbf{k}^i$ all decay exponentially with $e^{-kd/2}$ from the pattern boundary to the mid plane located at a distance $d/2$ from the pattern. Here, $k$ denotes the modulus of the primitive reciprocal unit vectors, $\xi$ is the correlation length of the magnetization fluctuations, $\delta M_z(\mathbf{r})$, which we assume to be the size of the cubes. Fluctuations in the cube magnetization $\delta M_z$ cause magnetic fields that also fulfill the Laplace equation but decay much slower from the pattern wall toward the mid-plane. The relative ratio of the relative fluctuation of the pattern field to the quasi-periodic pattern field amplitude is thus enhanced exponentially compared to the relative ratio of fluctuations of the magnetization $\delta M_z$ and the average cube magnetization $\bar{M}_z$.

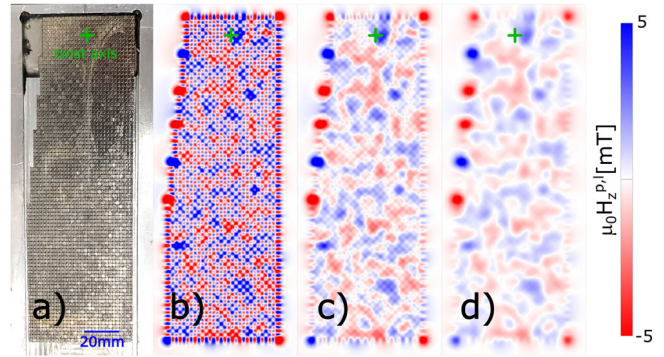

**Fig. 3 Effect of disorder on the pattern field. a** Image of the left pattern. The twist axis is marked as a green cross. Scale bar is 20 mm. **b**-**d** Measurement of the pattern field $H_z^{p,l}(\mathbf{r})$ of the pattern at the three elevations $z_a = 2$ mm, $z_b = 3$ mm, and $z_c = 4$ mm. The periodic part of the field associated with the antiferromagnetic ordering of the magnetic cubes decays faster with the elevation above the pattern than the disorder field arising from deviations of the saturation of the cube magnetization from the average value.

In Fig. 3a, we plot an image of one—namely, the left—of our two patterns. In Fig. 3b–d three measurements of the normal component of the magnetic field $H_z^{p,l}(\mathbf{r}, z)$ of the left of the two patterns at the three elevations $z_a = 2$ mm, $z_b = 3$ mm, and $z_c = 4$ mm are displayed. At the lowest elevation above the pattern, the field $H_z^{p,l}(\mathbf{r}, z_a)$ nicely reflects the antiferromagnetic arrangement of the individual cubes of the pattern. The scale bar runs from $-5$ mT to $5$ mT, reflecting that the field has dropped already significantly from the cubes saturation magnetization. Strong fields only occur at the corners of the pattern boundary where strong peaks in the field can be detected. Alternating cubes do not produce exactly the same magnetic field, and one can see positive patches where positively magnetized cubes dominate over the cubes of opposite magnetization; in other regions, there are negative patches. At medium elevations above the pattern, the field $H_z^{p,l}(\mathbf{r}, z_b)$ is a mixture of the pattern scheme with disordered field patches of equal strength. At an even larger distance $z_c$ to the pattern only the disordered patches remain.

**Trajectories in twisted patterns with disorder.** The disorder potential significantly changes the motion of a steel sphere through the midplane between the patterns. In Fig. 4, we plot overlays of both the experimental (**a**) and (**c**) and simulated (**b**) and (**d**) single particle trajectories of the steel sphere within the twisted patterns with magic twist angles (**a**) and (**b**): $\alpha_m(13) = 4.242°$, (**c**) and (**d**): $\alpha_m(7) = 7.628°$ and subject to an external field. The positions $\mathbf{r}(s)$ on the trajectories are parametrized by the arc length $s$. In consecutive experiments, the steel sphere enters the twisted pattern at different entry spots at the bottom of the pattern. The trajectories are colored (black to green) according to their flat channel parameter $f(\mathbf{r}(s))$ (for details, see the methods section). That is, the larger (greener), the longer the trajectories follow a direction of 45° to the upward direction. In the background, we show a color-coded plot of the quasi-periodic twisted potential with the superposed disorder contributions from both patterns. A magenta cross in Fig. 4a marks the axis of rotation.

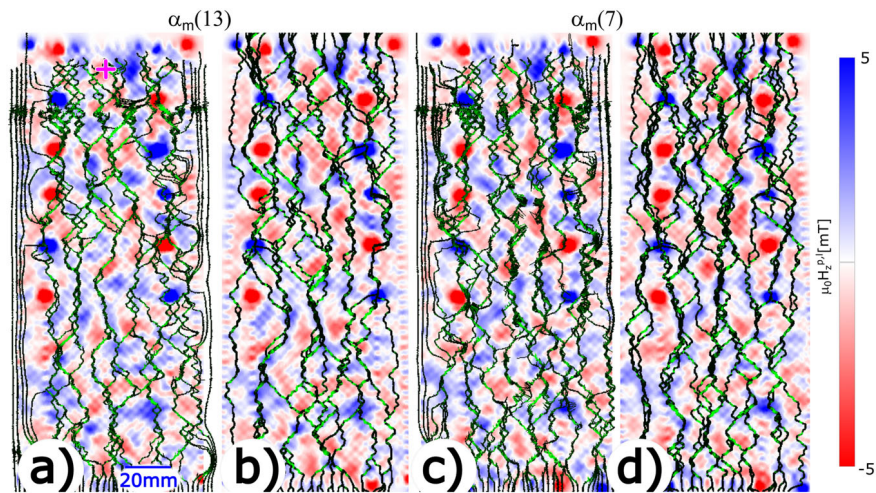

**Fig. 4 Comparison of experimental with simulated trajectories.** Trajectories of the steel spheres of diameter 1.9 mm within the twisted patterns with magic twist angles **a** experimental trajectories for $\alpha_m(13) = 4.242°$, **b** simulated trajectories for $\alpha_m(13) = 4.242°$, **c** experimental trajectories for $\alpha_m(7) = 7.628°$, and **d** simulated trajectories for $\alpha_m(7) = 7.628°$. The trajectories are colored according to their flat channel parameter. That is, the larger (greener), the longer a trajectory follows a direction of 45° (see also Methods Eq. (3)). The trajectories are plotted with the disordered quasi-periodic twisted potential as background. The potential was determined by the superposition of oppositely rotated single pattern potentials from the measurements shown in Fig. 3. A magenta cross in **a** marks the axis of rotation. Experimental trajectories beyond the left and the right borders of the pattern are straight lines, and the experimental trajectories inside the pattern stop at the air/water interface located at the elevation of the twist axis.

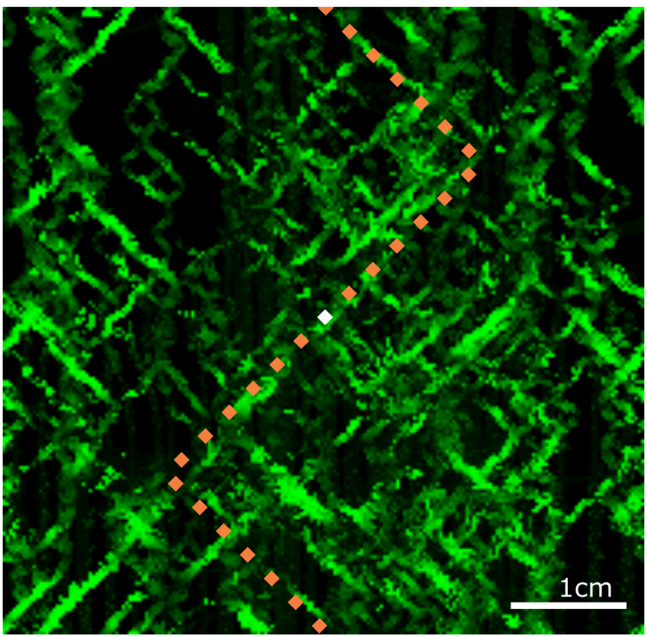

**Fig. 5 Flatness parameter of the measurement of Fig. 4a folded into the twisted Wigner Seitz cell $\mathcal{W}_{TW}$.** Green colors depict large flatness parameters, and low flatness parameters are depicted in black. The main flat channel runs along the orange dotted line. The axis of rotation is in the middle of the cell at the white spot.

In Fig. 5, we depict the maximum flatness parameter found in the trajectories of Fig. 4a and fold it into the twisted Wigner Seitz cell $\mathcal{W}_{TW}$. There is a significant correlation between the green parts of the trajectories where the flat channel parameter is large with the location of the main flat channel of the quasi-periodic potential (Eq. (1), orange dots in the Wigner Seitz cell $\mathcal{W}_{TW}$). Steel spheres, however almost never reach the corners of the twisted pattern, where the flat channel parameter is low. Therefore, there are many other subchannels that also show significant flatness within the Wigner Seitz cell other than the main channel.

Before the steel sphere reaches a corner, it is scattered into the interior away from the flat channels. This happens either by a red patch of the disorder potential in Fig. 4 blocking the flat channel or by a blue patch of the disorder potential luring the steel sphere into the mogul slope adjacent to the flat channel. These scatter processes are spaced at distances typical for the size of the random patches and are caused by the dispersion of magnetization amongst the magnetic cubes. Simulations of the steel sphere motion using the measured magnetic field of both patterns reproduce the measurements on the scale of the patches with small scale deviations on the scale of the size of the $\mathcal{P}$-Wigner Seitz cell.

**Anderson transition.** If we plot the mobility of particles in perfectly quasiperiodic twisted patterns[54] versus the twist angle, pronounced peaks in the mobility are found at the magic angles. The mobility peaks are caused by the easy passage of the steel sphere at the corners of the shift vectors of the two patterns at the rotation axis, which are arranged properly. For magic angles, the local shift vectors of the corners repeat at each corner of consecutive Wigner Seitz cells $\mathcal{W}_Q = \mathcal{W}_{TW}$ of the $\mathcal{Q}$-lattice. If we are not at a magic angle, the properties of consecutive corners that need to be passed are detuned from the easy-to-pass conditions because of the quasi-periodicity, and the motion becomes much harder to achieve. Since, for our system, the particles almost never reach the corners of the flat channels, the behavior of the trajectories at magic and at a nonmagic angle are no longer that different.

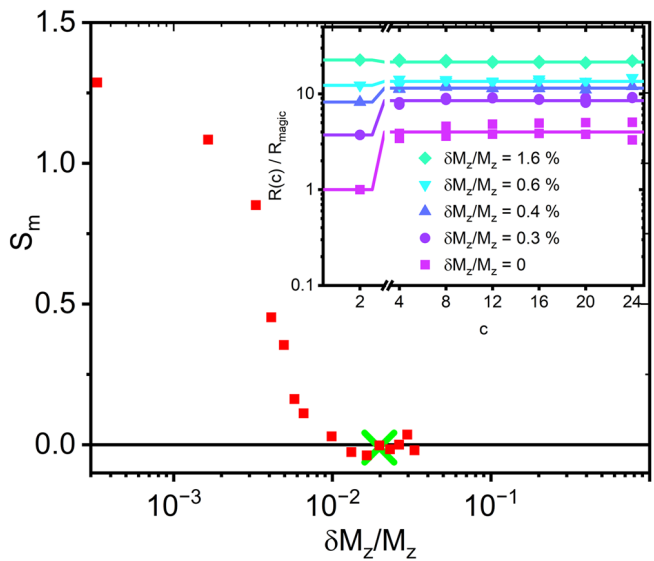

**Fig. 6 Disorder effects, Inset.** Simulated maximum resistance $R(c)/R_{magic}$ of the trajectories normalized to the magic resistance of a pure system. The resistance is plotted as a function of the number of corners on an ordered trajectory $c$. The larger $c$ the more non-magical the system becomes. The resistance for the magical $k = 7$ is much lower than all the other nonmagical angles nearby when the system is ordered but rises to the generic value for the larger disorder. *Main figure:* The simulated (red) and experimental (green data point) specificity of the system toward magic conditions $S_m = \ln(R(c \neq 2)/R(c = 2))$ as a function of the disorder $\delta M_z/M_z$. The disorder-induced destruction of magical specificity needs less than a percent of magnetization disorder.

One encounters $c = 2q2^{(p-q)\bmod 2} \in 2\mathbb{Z}$ flat channel corners on a trajectory of a commensurate twisted and disorder-free pattern within one twisted Wigner Seitz cell $\mathcal{W}_{TW}$. In the inset of Fig. 6 we plot the normalized maximum resistance $R = \max_s dr/ds \cdot \nabla U$ experienced by a simulated particle along the trajectory $r(s)$ versus the number of flat channel corners $c = 2q2^{(p-q)\bmod 2} \in 2\mathbb{Z}$. The normalization is made with the magic resistance $R_{magic}$ of a disorder free potential. The degree of non-magicness is a fluctuating function of the twist angle $\alpha$ but a systematic function of the number of different corners $c$. We therefore use nonmagical angles $\alpha(q_n = n, p_n = n(2k+1)\pm1, 2)$ close to the magic angle $\alpha_m(k)$ with the least amount of corners possible. The larger is $n$, the larger is $c$, and the less magic is the character of the twist angle. The larger is $n$, the closer is the angle $\alpha(q_n, p_n)$ to the magical angle $\alpha_m(k)$. The series of resistances is shown for $k = 7$. In an ordered twisted pattern, the resistance is lowest for magic ($c = 2$) angles and sharply rises with other $c$-values ($c \neq 2$). In contrast to this, in a strongly disordered system, the resistance of a magic ($c = 2$) angle no longer differs from other ($c \neq 2$) values. The logarithm of the magic ratio $S_m = \ln(\frac{R(c \neq 2)}{R(c = 2)})$ is, therefore, a measure of the specificity of the system at magic angles. In the main part of Fig. 6 we plot the simulated magic specificity of the twisted system as a function of the magnetization disorder $\delta M_z/M_z$. The scatter of our magnetic cubes in the experiment was $\delta M_z/M_z \approx 2\%$. The experimental specificity is computed using the measured potential (at the fixed scatter of the experimental magnetic cube magnetization) and the measured experimental trajectories. The magic specificity of the system decreases with the disorder of the lattice until it drops to the magic nonspecific value of $S_m = 0$. The specificity of the magic behavior thus disappears with increasing disorder. This lets us conclude that fluctuations and already weak magnetization disorder have a pronounced effect on the transport behavior in twisted quasi-periodic systems.

## Discussion

It would be nice to experimentally prove the Anderson transition simulated in this paper. Systematic experiments on the Anderson transition would require the use of different sets of magnetic cube patterns consisting of more precise cubes with less scatter of their magnetization. From Fig. 6, we expect that experiments with one order of magnitude more precise cubes would show the return of the trajectories into the flat channels and the Anderson back transition from generic resistance, as observed experimentally in this paper, toward magic resistance peaks as predicted in reference[54]. An alternative approach could be to systematically reduce the separation of the two patterns without losing the simplicity of the periodic part of the potential that is only a sum of longest wave Fourier modes when the midplane between both patterns is at sufficient distance to the pattern. This would require sinusoidally magnetized magnetic cubes.

In electronic transport measurements of graphene, each carbon atom exactly resembles the other such that the randomness of the building blocks cannot be used in the same way as in this work to induce similar kinds of transitions. One generic form of randomness in twisted bilayer graphene systems is unintentional as well as intentional heterostrain. In intentionally heterostrained twisted bilayer graphene[55], the heterostrain renormalizes the flat bands in a way that strongly depends on experimental details and transport properties are consequently altered. Van Hove singularities can give rise to flat energy bands that are or are not connected at the corners in the Brillouin zone. The corners amount to a tiny fraction of the area both in direct and in reciprocal space. We believe that this is part of the reason that transport properties are fragile and can be altered easily.

## Conclusions

The transport behavior of steel spheres between twisted macroscopic antiferromagnetic square patterns in a superposed external magnetic field shows a pronounced drop of the resistance if the twist angles are magic and the system is free of disorder[54]. Disorder scatters particles into the mogul slope potential adjacent to flat channels, and an Anderson transition lets the magic transport behavior return to generic nonmagic high resistance behavior. The system, therefore, shares similarities with the behavior of electrons between twisted monolayers.

## Methods

**Twist angle calibration and measurement**. We determine the zero twist angle position by placing both patterns with no water tank in between on top of each other such that the cubes of one and the other pattern are separated by less than 0.5 mm. The magnetic interaction between both patterns is strong enough to enforce a perfect twist-free alignment ($\alpha < 1/100 n_y$ with $n_y = 87$, the number of cubes along the long axis of the pattern). Two laser pointers are mounted to the backside of one and the other pattern. The lasers are adjusted to point to the same spot of a plane at a distance of $l = 1.8\,m$ from the twist axis. When we adjust a certain twist angle after the water tank was placed between the patterns we can read the twist angle with a precision of $\delta\alpha \approx \tan\delta\alpha = \delta s/l \approx 0.5 \times 10^{-4}$ from the separation $s$ of the two laser spots at the plane.

**Overdamped dynamics simulations**. We simulate the system using overdamped Brownian dynamics in the limit $T \to 0$. The equation of motion is integrated in time using the standard Euler algorithm. The steel sphere is modeled as a point particle.

**Flat channel parameter**. We compute the flat band parameter as

$$f(\boldsymbol{r}(s)) = \max_{L < a^c} \left[ \sum_{\{s - L, s, s + L\} s_i > s_j \in} L^c \times \right.$$

$$\left. \times \exp\left\{ -\left[\frac{\boldsymbol{e}_z \cdot \frac{\boldsymbol{r}(s_i) - \boldsymbol{r}(s_j)}{|\boldsymbol{r}(s_i) - \boldsymbol{r}(s_j)|} - \frac{1}{\sqrt{2}}}{\Delta^2}\right]^2 \right\} \right]$$

(3)

with $\epsilon_{exp} = \epsilon_{sim} = 3$ and $\Delta^2_{exp} = 0.0005$, $\Delta^2_{sim} = 0.0055$ for the experiments (simulations). The Gaussian in Eq. (3) ensures that all sides of the triangle $\boldsymbol{r}(s)$, $\boldsymbol{r}(s + L)$, and $\boldsymbol{r}(s - L)$ are almost degenerate to a line and all sides enclose an angle of nearly $\pi/4$ with $\boldsymbol{e}_z$. The factor $L^c$ increases the flat channel parameter the larger the triangle that fulfills this condition.

## Data availability

All the data supporting the findings are available from the corresponding author upon reasonable request.

## Acknowledgements

We thank Daniel de las Heras for bringing the problem of transport in twisted magnetic patterns to our attention, Oliver Bäumchen and Reinhard Richter for letting us use their pair of Helmholtz coils. We acknowledge funding by the Deutsche Forschungsgemeinschaft (DFG, German Research Foundation) under project number 440764520.

## Author contributions
A.R. designed and performed the experiments, A.E. performed the simulations, A.R. and T.M.F. wrote the paper, and A.R. and M.D. built the setup.

## Funding
Open Access funding enabled and organized by Projekt DEAL.

## Competing interests
The authors declare no competing interests

## Additional information
**Supplementary information** The online version contains supplementary material available at https://doi.org/10.1038/s42005-023-01512-6.

**Correspondence** and requests for materials should be addressed to Thomas M. Fischer.

**Peer review information** *Communications Physics* thanks the anonymous reviewers for their contribution to the peer review of this work. A peer review file is available.

**Reprints and permission information** is available at http://www.nature.com/reprints

**Publisher's note** Springer Nature remains neutral with regard to jurisdictional claims in published maps and institutional affiliations.